\documentclass{PoS}

\usepackage{graphicx}
\usepackage{xspace}
\usepackage{floatrow}
\usepackage{amsmath}
\usepackage{amsfonts}
\usepackage{amssymb}
\usepackage{calrsfs}
\usepackage{wrapfig}
\usepackage{multirow}
\usepackage{color}

\newcommand{\epem}{\rm e^+e^-}
\newcommand{\alphas}{\alpha_{\rm s}}
\newcommand{\alphasmZ}{\alphas(\rm m^2_{_{\rm Z}})}
\newcommand{\alphasnew}{\alphas^{^{\rm new}}(\rm m^2_{_{\rm Z}})}
\newcommand{\sqrts}{\sqrt{s}}
\newcommand{\qqbar}{q\overline{q}}
\newcommand{\ccbar}{c\overline{c}}
\newcommand{\bbbar}{b\overline{b}}
\newcommand{\ttbar}{t\overline{t}}

\newcommand{\gaga}{\gamma\,\gamma}

\newcommand{\Rlz}   {R^0_\ell}

\newcommand{\Ghad}  {\rm\Gamma_{\mathrm{had}}}

\newcommand{\GZ}    {\rm \Gamma_{\mathrm{Z}}}
\newcommand {\so}   {\rm \sigma_0^{had}}

\newcommand*{\eg}{e.g.\@\xspace}
\newcommand*{\ie}{i.e.\@\xspace}

\def\ttt#1{\texttt{\scriptsize #1}}

\newfloatcommand{capbtabbox}{table}[][\FBwidth]

\title{$\alphas$ status and perspectives (2018)}
\ShortTitle{$\alphas$ status and perspectives (2018)}

\author{\speaker{David d'Enterria}\\ 
       CERN, EP Department, 1211 Geneva, Switzerland\\
       E-mail: \email{dde@cern.ch}}

\abstract{
The latest experimental and theoretical developments in the high-precision determination 
of the strong coupling $\alphas$ are briefly reviewed. 
Six groups of observables: (i)~lattice QCD data, (ii)~hadronic $\tau$ decays, 
(iii)~deep-inelastic e$^\pm$p data and parton distribution functions (PDF) fits, 
(iv)~event shapes and jet rates in $\epem$ collisions, 
(v)~Z boson hadronic decays, and (vi)~top-quark cross sections in pp collisions,
are used to extract the current world-average at the Z pole mass, $\alphasmZ = 0.1181~\pm~0.0011$
at next-to-next-to-leading-order (NNLO), or beyond, accuracy.
Additional NNLO extractions have recently appeared based on new lattice studies, the $R(s)$ ratio 
in $\epem\to \mbox{hadrons}$, updated PDF fits, energy-energy correlations in $\epem$ collisions, 
jet cross sections in e$^\pm$p collisions, and the full set of pp\,$\to\ttbar$ cross sections at the LHC. 
Inclusion of these new data into the world-average would slightly increase its value and reduce its
uncertainty to $\alphasmZ = 0.1183~\pm~0.0008$.
Future $\alphas$ extraction perspectives with permille uncertainties at future high-luminosity 
$\epem$ machines -- via W and Z hadronic decays, parton fragmentation functions, and photon
F$_2$ structure function in $\gamma\,\gamma$ collisions -- are also discussed.}

\FullConference{XXVI International Workshop on Deep-Inelastic Scattering and Related Subjects (DIS2018)\\
		16-20 April 2018\\
		Kobe, Japan}

\begin{document}

\section{Introduction}

The Lagrangian of the theory of the strong interaction, Quantum Chromodynamics (QCD), has, besides the 
quark masses, a single free parameter: the $\alphas$ coupling that determines the strength of the interaction 
between quarks and gluons at a given energy~\cite{SalamBethke,d'Enterria:2015toz,Deur:2016tte}. 
Due to its logarithmic decrease with energy (asymptotic freedom), $\alphas$ is commonly 
given at the reference scale of the Z pole mass. Its current value, $\alphasmZ = 0.1181~\pm~0.0011$~\cite{PDG}, 
has a $\delta\alphas/\alphas \approx 0.9\%$ uncertainty 
that is orders of magnitude worse than that of the electromagnetic, Fermi, and gravitational couplings: 
$\delta\alpha/\alpha \approx 10^{-10} \ll \delta G_{\rm F}/G_{\rm F} \approx 10^{-8} \ll \delta G/G \approx 10^{-5}$.
The strong coupling is one of the fundamental parameters of the Standard Model (SM), and its value 
affects chiefly all theoretical calculations of perturbative QCD (pQCD) processes 
involving partons, leading \eg\ to 3--7\% uncertainties in key Higgs processes  such as $gg\to H$ and 
associated $H$-$\ttbar$ cross sections, and $H\to\bbbar,\ccbar,gg$ branching fractions. 
In addition, the QCD coupling precision dominates the parametric uncertainties in future determinations of the 
top mass~\cite{Hoang:2017suc} and electroweak precision observables~\cite{TLEP}. Last but not least,
$\alphas$ also impacts physics approaching the Planck scale, either 
in the electroweak vacuum stability~\cite{Buttazzo:2013uya} or in searches of new coloured 
sectors that may modify its running towards the GUT scale~\cite{d'Enterria:2015toz}. 

\section{Current $\alphasmZ$ world average and updates}
\label{sec:current}

The current world-average $\alphasmZ$ is based on the combination of six subclasses of 
(appro-ximately-independent) observables measured at various energies~\cite{PDG} that are listed in Table~\ref{tab:alphas}.
Each extraction is summarized below, with newly derived values not included in the 
current PDG-2017 world-average quoted as \textcolor{red}{$\alphasnew$}:\vspace{0.3cm}\\
{\bf (1)} The comparison of NNLO pQCD predictions to computational {\bf lattice QCD} results (Wilson loops, 
$\qqbar$ potentials, hadronic vacuum polarization, QCD static energy) constrained by the experimental 
hadron masses and decay constants, provides the most precise $\alphas$ extraction: $\alphasmZ = 0.1188 \pm 0.0011$
with a~0.9\% uncertainty dominated by finite lattice spacing, pQCD expansion truncations, and hadron extrapolations. 
A new analysis of the ALPHA collaboration with reduced pQCD uncertainties reports \textcolor{red}{$\alphasnew = 0.11852 \pm 0.00084$}
(0.7\% uncertainty)~\cite{Bruno:2017gxd}.
Further reduction of the statistical uncertainties, at least by a factor of two, can be anticipated
with increased computing power over the next 10 years.\vspace{0.3cm}\\
{\bf (2)} The ratio of hadronic to leptonic {\bf tau decays}, known experimentally to within $\pm0.23\%$ 
($R_{\rm \tau,exp} = 3.4697 \pm 0.0080$), compared to next-to-NNLO (N$^3$LO) calculations, 
yields $\alphasmZ = 0.1192 \pm 0.0018$, \ie\ 
a 1.5\% uncertainty, through a combination of results from different pQCD approaches 
(CIPT and FOPT, with different treatments of non-pQCD corrections)~\cite{Pich:2016bdg,Boito:2016oam}. 
Applying the same calculational techniques for the measured $R(s)$ ratio in $\epem \to$\,hadrons for 
$\sqrts < 2$\,GeV, a value \textcolor{red}{$\alphasnew = 0.1162 \pm 0.0025$} ($\sim$2\% uncertainty) has been recently derived~\cite{Boito:2016oam}. 
\vspace{0.3cm}\\
{\bf (3)} A combination of various analyses of {\bf deep-inelastic scattering} (including N$^3$LO fits of 
$F_{2}(x,Q^2)$, $F^c_2(x,Q^2)$, and $F_L(x,Q^2)$) and {\bf global PDF fits} yield a central value $\alphasmZ = 0.1156 \pm 0.0021$ 
with 
1.8\% precision. An updated NNLO PDF fit from the NNPDF3.0 collaboration reports \textcolor{red}{$\alphasnew = 0.1185 \pm 0.0012$}
($\sim$1\% uncertainty)~\cite{Ball:2018iqk}. 
Also, jet cross section data from e$^\pm$p collisions at HERA~\cite{Andreev:2017vxu}, compared to state-of-the-art
NNLO calculations~\cite{NNLO_ep_jets}, yield \textcolor{red}{$\alphasnew = 0.1157 \pm 0.0035$} ($\sim$3\% uncertainty).
Ultimate uncertainties of order 
0.3\% require similar measurements at a future high-luminosity DIS machine (such as LHeC or FCC-eh)~\cite{lhec}.\vspace{0.3cm}\\
{\bf (4)} LEP measurements of {\bf $\mathbf{\epem}$ event shapes and jet rates} (thrust, C-parameter, N-jet cross sections) analysed with 
N$^{2,3}$LO calculations matched, in some cases, with soft and collinear resummations at N$^{(2)}$LL accuracy,
yield $\alphasmZ$~=~0.1169~$\pm$~0.0034, with a 
2.9\% uncertainty mostly driven by the span of individual extractions which use different (Monte Carlo or analytical) 
approaches to correct for hadronization effects.
Modern jet substructure techniques~\cite{Baron:2018nfz} can help mitigate the latter corrections.
A recent NNLO+NNLL analysis of energy-energy correlations in $\epem$ yields
\textcolor{red}{$\alphasnew = 0.1175 \pm 0.0029$} ($\sim$2.5\% uncertainty)~\cite{Kardos:2018kqj}.
\vspace{0.3cm}\\
{\bf (5)} Three {\bf hadronic Z decay} observables measured at LEP ($\GZ$, $\so = 12 \pi/m_Z \cdot \Gamma_e\Ghad/\Gamma_Z^2$, and
$\Rlz = \Ghad/\Gamma_\ell$) compared to N$^3$LO calculations, yield 
$\alphasmZ = 0.1196 \pm 0.0030$ 
with 2.5\% uncertainty. Fixing all SM parameters to their
measured values and letting free $\alphas$ in the latest electroweak fit yields \textcolor{red}{$\alphasnew = 0.1194 \pm 0.0029$}
($\sim$2.4\% uncertainty)~\cite{Haller:2018nnx}. Permille level precision will require large-statistics measurements accessible \eg\ with 
10$^{12}$ Z bosons at the FCC-ee~\cite{TLEP}.\vspace{0.3cm}\\
{\bf (6)} Theoretically known at NNLO+NNLL, 
top-pair cross sections are the first {\bf hadron collider}
measurements that constrain $\alphas$ at this level of accuracy. The PDG-2017 contains a single extraction from the CMS data, 
$\alphasmZ = 0.1151 \pm 0.0028$, with a 
2.5\% uncertainty mostly dominated by the gluon PDF uncertainties~\cite{Chatrchyan:2013haa}. A recent combination 
of all $\ttbar$ LHC and Tevatron data increases its value to \textcolor{red}{$\alphasnew = 0.1177 \pm 0.0035$}, with a larger ($\sim$3\%) 
uncertainty~\cite{Klijnsma:2017eqp}. Novel jet cross sections calculations at NNLO~\cite{Currie:2016bfm} 
will allow to fully exploit the multiple jet datasets available 
for additional upcoming precise $\alphas$ extractions~\cite{Britzger:2017maj}.
\begin{table}[htbp!]
\renewcommand{\arraystretch}{1.15}
\caption{\label{tab:alphas} QCD coupling PDG-2017~\cite{PDG} values extracted in six subclasses of observables, 
associated pre-averages, and world-average. 
New $\alphasmZ$ extractions and recomputed averages are listed in red italics.}%
\begin{tabular}{lc@{}@{}c@{}@{}c}\hline\hline
Observable class       & $\alphasmZ$ extractions \textcolor{red}{({\it new})} & $\alphasmZ$ average \textcolor{red}{({\it new})} \\\hline  
lattice QCD            &  0.1184(6), 0.1192(11), 0.1182(7), 0.1205$^{+0.0009}_{-0.0019}$, & 0.1188(11)  & \\
                       &  0.1196(12), 0.1166$^{+0.0012}_{-0.0008}$, \textcolor{red}{{\it 0.11852(84)}} & \textcolor{red}{{\it 0.1187(10)}}\\\hline  
hadronic $\tau$ decays &  0.1202(19), 0.1200(15), 0.1199(15), & 0.1192(18) \\
                       &  0.1165(19), 0.1193(23), \textcolor{red}{{\it 0.1162(25)}} & \textcolor{red}{{\it 0.1187(19)}}\\\hline  
DIS and PDF fits       &  0.1134(25), 0.1141(22), 0.1158(36)  & 0.1156(21) \\
                       &  0.1172(13), 0.1173(11)$\to$\textcolor{red}{{\it 0.1185(12); 0.1157(35)}}  & \textcolor{red}{{\it 0.1158(24)}}\\\hline  
$\epem$ shapes,        &  0.1224(39), 0.1189(43), 0.1172(51),  & 0.1169(34) \\
and jet rates          &  0.1175(25), 0.1164$^{+0.0028}_{-0.0024}$, 0.1137$^{+0.0034}_{-0.0027}$, & \textcolor{red}{{\it 0.1169(33)}}\\ 
                       &  0.1135(11), 0.1123(15), 0.1199(59),  \textcolor{red}{{\it 0.1175(29)}} & \\\hline  
hadronic Z decays      &  0.1196(30); \textcolor{red}{{\it 0.1194(29)}}  & 0.1196(30); \textcolor{red}{{\it 0.1194(29)}} \\\hline  
pp$\to\ttbar$ cross sections &  0.1151(28); \textcolor{red}{{\it 0.1177(35)}} & 0.1151(28); \textcolor{red}{{\it 0.1177(35)}} \\\hline  
world average                &    & 0.1181(11); \textcolor{red}{{\it 0.1183(8)}} \\\hline\hline
\end{tabular}
\end{table}

\begin{figure}[htpb!]
\includegraphics[width=0.65\linewidth]{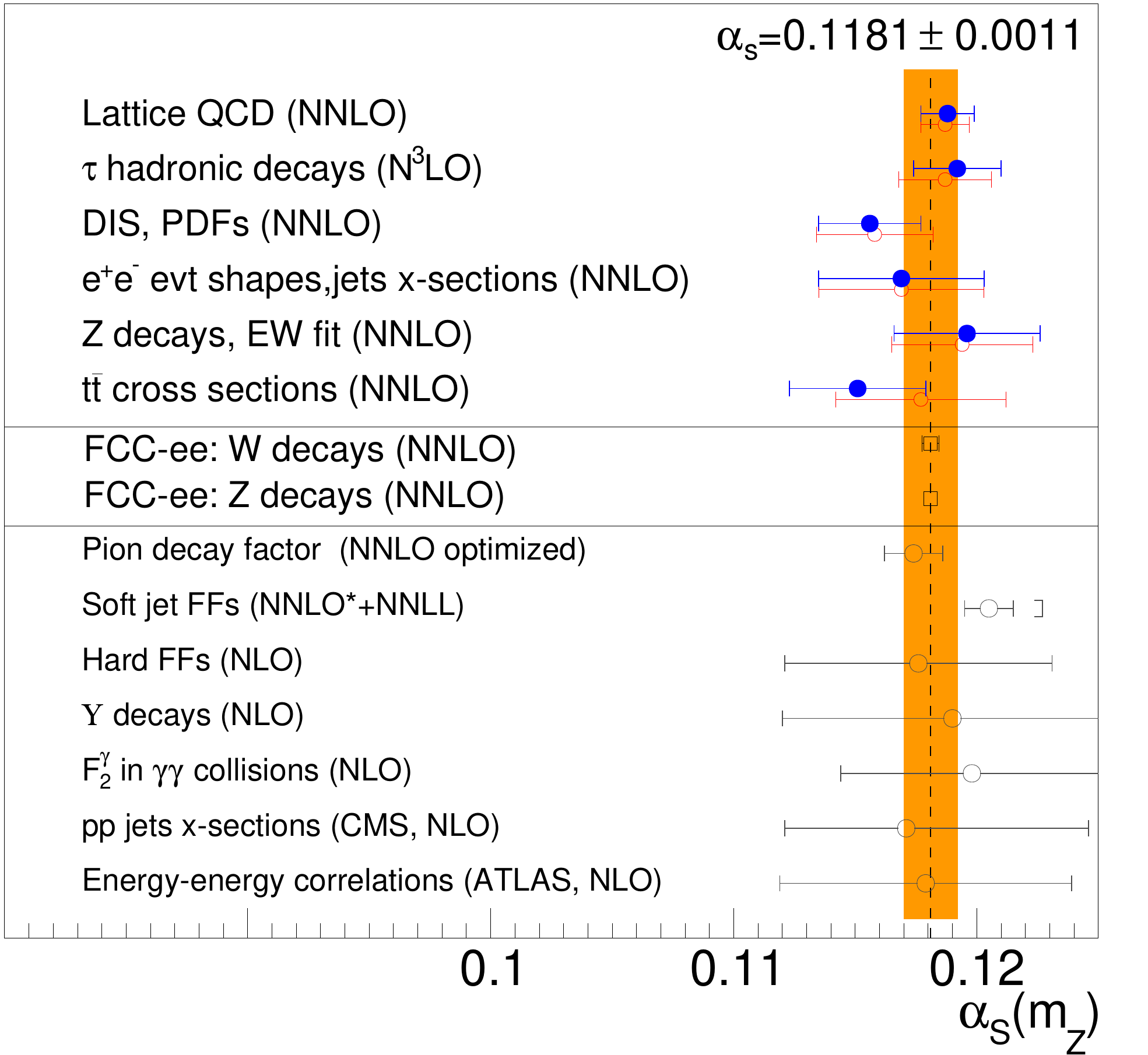}
\caption[]{$\alphas$ extractions. Top: Current PDG-2017 (solid dots, orange band) and 2018-updated (open dots) pre-averages.
Middle:\,Expected FCC-ee values via W, Z decays.\,Bottom:\,Other less accurate methods today.
}
\label{fig:alphas}
\end{figure}

\noindent Table~\ref{tab:alphas} summarizes all high-precision $\alphas$ values extracted so far. The $\chi^2$-averaging of the 
six subgroups of observables currently in the PDG-2017 yields $\alphasmZ = 0.1181~\pm~0.0011$~\cite{PDG}. 
Inclusion of the newly derived (red-italics) values has 
almost no impact in four subclasses (lattice QCD, PDF, $\epem$, Z decays) but would change by $-0.4\%$ ($+2\%$)
the $\tau$- (top)-based pre-averages (Fig.~\ref{fig:alphas}). The updated world-average, combining all results,
would thereby be $\alphasmZ = 0.1183~\pm~0.0008$ with slightly increased central value and decreased uncertainty ($\sim$0.7\%).\\

\section{Future $\alphas$ prospects}
\label{sec:others}

Improvements in a few extractions listed in Table~\ref{tab:alphas} are anticipated in the coming
years thanks to new LHC data and more precise calculations. In addition, other sets of observables
computed today with a lower accuracy (NLO, or approximately-NNLO, bottom of Fig.~\ref{fig:alphas}), 
and thereby not included now in the world-average, will provide additional constraints~\cite{d'Enterria:2015toz}.
Ultimately, $\alphasmZ$ precision in the permille range will require a clean $\epem$ machine providing many 
orders-of-magnitude more jets and electroweak bosons than collected at LEP. 
Measurements of {\bf W hadronic decays} (theoretically known at N$^3$LO) 
provide today a very imprecise $\alphasmZ$ = 0.117~$\pm$~0.030 ($\sim$30\% uncertainty) due to the limited 
LEP data. Statistical samples of 10$^8$ W available at FCC-ee~\cite{TLEP}, combined with a 
significantly reduced parametric uncertainty of the $V_{cs}$ CKM element, 
can ultimately yield $\delta\alphasmZ/\alphasmZ\approx$~0.3\%~\cite{dEnterria:2016rbf}.
Similarly, the high-statistics and clean set of accurately-reconstructed (and flavour-tagged) $\epem$ final-states 
will provide precise $\alphas$ determinations from event shapes, jets rates,
and parton-to-hadron {\bf fragmentation functions (FF)} studies. The energy dependence of the low-$z$ FF provides today
$\alphasmZ = 0.1205 \pm 0.0022$ ($\sim$2\% uncertainty) at NNLO*+NNLL~\cite{softFF}, whereas
NLO scaling violations of the high-$z$ FFs yield $\alphasmZ = 0.1176 \pm 0.0055$ ($\sim$5\% uncertainty, 
mostly of experimental origin)~\cite{Albino:2005me}. 
Also, measurements of the {\bf photon structure function} F$_2^\gamma(x,Q^2)$, via $\epem \to \gaga\to$\,hadrons, have been 
used to obtain $\alphasmZ$ = 0.1198~$\pm$~0.0054 ($\sim$4.5\% uncertainty) at NLO~\cite{Albino:2002ck}. 
Extension to full-NNLO accuracy of the FF and F$_2^\gamma(x,Q^2)$ fits using much larger $\epem$ datasets 
available at various center-of-mass energies at FCC-ee would allow reaching 
subpercent precision in those $\alphas$ extractions.\\ 


\noindent {\bf Acknowledgments} I am grateful to S.~Bethke and G.~Salam for useful discussions, and to 
R.~P\'erez-Ramos and M.~Srebre for common work leading to a couple of new $\alphas$ extractions reported here.


\end{document}